\documentclass[aps,twocolumn,pre,showkeys,eqsecnum]{revtex4}
\usepackage{graphpap}
\usepackage[dvips]{graphicx}
\usepackage[dvips]{graphics}
\usepackage{color}

\begin{document}

\title{Electronic Excited States in Bilayer Graphene Double Quantum Dots}
 \author{C. Volk$^{1,2}$, S. Fringes$^{1}$, B. Terr\'es$^{1,2}$, J. Dauber$^{1}$, S. Engels$^{1}$, S. Trellenkamp$^2$, and C. Stampfer$^{1,2}$}

 \affiliation{
$^1$JARA-FIT and II. Institute of Physics B, RWTH Aachen University, 52074 Aachen, Germany\\
$^2$Peter Gr\"unberg Institute (PGI-8/9), Forschungszentrum J\"ulich, 52425 J\"ulich, Germany
}

\date{ \today}

 \begin{abstract}
We report tunneling spectroscopy experiments on a bilayer graphene double quantum dot device that can be tuned by all-graphene lateral gates. The diameter of the two quantum dots are around 50~nm and the constrictions acting as tunneling barriers are 30 nm in width. The double quantum dot features addition energies on the order of 20~meV. Charge stability diagrams allow us to study the tunable interdot coupling energy as well as the spectrum of the electronic excited states on a number of individual triple points over a large energy range. The obtained constant level spacing of $1.75$~meV over a wide energy range is in good agreement with the expected single-particle energy spacing in bilayer graphene quantum dots. Finally, we investigate the evolution of the electronic excited states in a parallel magnetic field.

 \end{abstract}


% \pacs{71.10.Pm, 73.21.-b, 81.05.Uw, 81.07.Ta}
 \maketitle

\newpage

Graphene quantum dots (QDs) are interesting candidates for
spin qubits with long coherence times~\cite{tra07}.
The suppressed hyperfine interaction and weak spin-orbit
coupling~\cite{min06,hue06} make graphene and flat carbon structures in general,
promising for future quantum information technology~\cite{los98}.
Significant progress has been made recently in the fabrication
and understanding of graphene quantum devices.
A "paper-cutting" technique enables the fabrication of
graphene nanoribbons~\cite{che07,han07,mol09,sta09,tod09,liu09,gal10,han10,ter11},
quantum dots~\cite{sta08b,pon08,sch08,gue09,mos10,gue10}, and double quantum dot devices~\cite{mol09a,mol10a,liu10,wan10}, where
a disorder-induced energy gap allows confinement of individual carriers in graphene.
These devices allowed the experimental investigation of
 excited states~\cite{sch08,mol10a}, spin states~\cite{gue10} and
the electron-hole crossover~\cite{gue09}.
However, all of these studies were based on single-layer graphene
and showed a number of device limitations related to
the presence of disorder, vibrational excitations and to the fact
that the missing band gap makes it difficult to realize
soft confinement potentials and "well-behaving" tunneling
barriers. In particular, it has been shown that intrinsic
ripples and corrugations in single-layer graphene can
lead to unintended vibrational degrees of freedom~\cite{mas10}
and to a coherent electron-vibron coupling in graphene QDs~\cite{rou11}.
Bilayer graphene is a promising candidate to overcome some of these limitations.
In particular it allows to open a band gap by an out-of-plane electric field~\cite{cas07a,oos08,zha09}, which may enable a
soft confinement potential and may reduce the influence of localized edge states. More importantly, it has been shown that ripples and substrate-induced
disorder are reduced in bilayer graphene~\cite{mey07a}, which increases the mechanical stability and suppresses unwanted vibrational modes.
\begin{figure}[t]\centering
\includegraphics[draft=false,keepaspectratio=true,clip,%
                   width=1.0\linewidth]%
                   {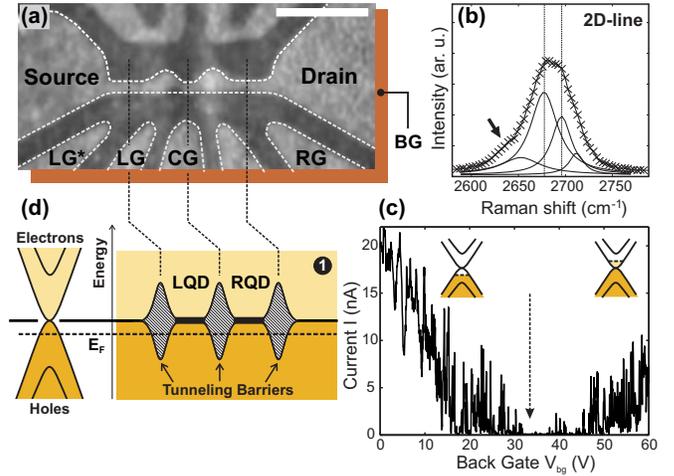}
\caption[FIG1]{(color online)
(a) Scanning force microscope image of the investigated double quantum dot device. The scale bar is 200~nm.
(b) 2D-line of the Raman spectrum fitted with four Lorentzians. The separation of the inner peaks measures 18.8~cm$^{-1}$.
(c) Back gate characteristics recorded at $V_b = 20~$mV and $T = 1.3~$K. The arrow marks the BG regime where the following measurements have been taken.
(d) Schematic illustration of the effective band structure of the device highlighting the three tunneling barriers (hatched areas) induced by the local constrictions (see panel a).

}
\label{fig1}
\end{figure}

Here, we present a bilayer graphene double quantum dot (DQD) device, with
a number of lateral gates. These local gates
allow to tune transport from hole to electron
dominated regimes and they enable to
access different device
configurations. We focus on the DQD configuration and
show characteristic honeycomb-like charge stability diagrams, with addition energies on the order of
20~meV. Most interestingly, we observe electronic excited states, which
are equally spaced over a wide energy range, in
good agreement with the expected single-particle confinement energy in bilayer
graphene QDs.

The DQD device shown in Figure 1a is fabricated based on bilayer graphene obtained from mechanical exfoliation of bulk graphite. The bilayer graphene flakes are deposited on a highly p-doped silicon substrate with a 295~nm silicon oxide layer. Raman spectroscopy measurements are used to unambiguously identify bilayer graphene. In Figure 1b we show the 2D Raman peak recorded on the flake which has been used to fabricate the device shown in Figure 1a. In contrast to single-layer graphene we find that four Lorentzians are required for fitting the 2D line shape. In particular, the left shoulder (see arrow in Fig. 1b) and the spacing between the inner two peaks, which is 18.8~cm$^{-1}$ (see dashed lines in Fig. 1b) provide a clear fingerprint for bilayer  graphene~\cite{gra07,fer06}. Electron beam (e-beam) lithography is used to pattern the device etch mask on the isolated graphene flake. Reactive ion etching based on Ar/O$_2$ plasma is employed to transfer the pattern to the bilayer graphene. Finally, an additional e-beam and lift-off step is used to contact the nanostructure with Cr/Au (5~nm/50~nm) electrodes.

Figure 1a shows a scanning force micrograph of the investigated bilayer DQD. The diameters of the etched quantum dots (QDs) measures roughly 50~nm while the width of the 100 nm long constrictions leading to the dots measure 30~nm.
For tuning the individual tunneling barriers and QDs, the device includes lateral bilayer graphene gates positioned in a distance less than 80~nm next to the active structure.
In this study we used the left and right gate (LG and RG) to change the number of
carriers in the left and right QD (LQD and RQD), respectively, while the central gate (CG) and the second left gate (LG*) are used
to tune the inter-dot coupling and the left barrier, respectively.
Additionally, the back gate (BG) is used to adjust the overall Fermi level.

The measurements have been performed in a pumped $^4$He system with a base temperature of 1.3~K and in a dilution refrigerator with an electron temperature around 100~mK. We have measured the two-terminal conductance through the bilayer graphene DQD by applying a symmetric dc bias voltage $V_b$ while measuring the current through the device with a resolution better than 50~fA.

The source-drain current of the device measured at finite bias ($V_b = 20~$mV) over
a large BG voltage range is shown in Figure 1c.
In close analogy to single-layer graphene nanodevices we observe a region of suppressed
current separating the hole (left inset) from the electron transport regime (right inset).
This so-called transport gap~\cite{sta09}, which extends for the investigated device from roughly
32~V to 48~V is expected to be mainly caused by the local tunneling barriers formed by
the three 30~nm narrow constrictions (see illustration in Figure 1d)~\cite{sta08b,gue09}.
The reproducible sharp conductance resonances in and around the transport gap region are due to localized states in
the constrictions whereas the overall hole-doping is most likely
due to atmospheric O$_2$ binding~\cite{ryn10}.

In order to access the DQD regime we fix the BG voltage to
a value inside the transport gap, such that the source and drain potentials are within the valence band
(see arrow in Figure 1c and Fermi level depicted
in Figure 1d).
\begin{figure}[tb]\centering
\includegraphics[draft=false,keepaspectratio=true,clip,%
                   width=1.0\linewidth]%
                   {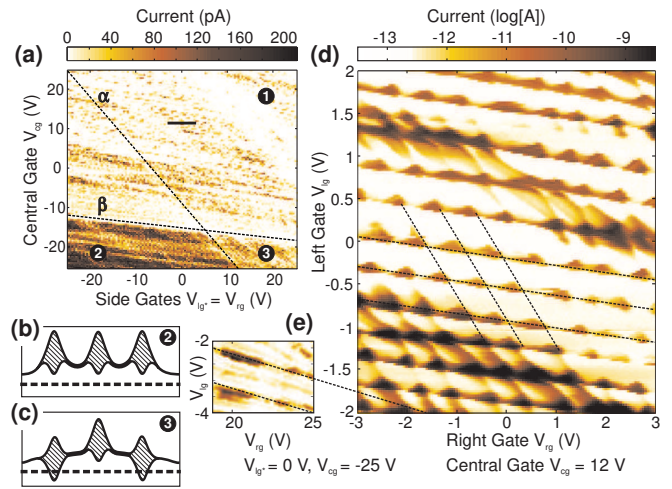}
\caption[FIG2]{
(color online) (a) Source-drain current as a function of the central gate voltage ($V_{cg}$) and the voltages applied to the outer side gates ($V_{lg^*}$=$V_{rg}$) at $V_b = 20~$mV and $T = 1.3~$K.
(b,c) Schematic band structure depicting the pure hole transport regime (2) and a single quantum dot regime with lifted central barrier (3), respectively. The dashed lines denote the position of the Fermi level.
(d) Charge stability diagram at finite bias ($V_b = 10~$mV) recorded in the double quantum dot regime  (regime (1), see also Figure 1d) as highlighted by the black bar in panel (a). (e) Local resonances in the right constriction as function of the left and right gate voltages measured in regime (3) with additionally lifted left tunneling barrier ($V_{lg^*}=0$ and $V_{lg}<0$), such that only the right tunneling barrier crosses the Fermi level ($V_b = 10~$mV)~\cite{com01a}.
}
\label{fig2}
\end{figure}
In Figure 2a we show a measurement of the current as function of the
central gate voltage and the left and right gate voltages ($V_{lg^*}=V_{rg}$) while keeping the BG voltage fixed (arrow in Fig. 1c).
Two characteristic slopes are observed which
separate areas of suppressed current from the area of elevated
current in the lower left corner.
This measurement suggests that the energy diagram shown in Figure 1d (including
a dominant central barrier) is a useful description for the three different transport regimes (1)-(3) found
in the data.
In regime (2) all three side gates are tuned to very negative voltages, such
that all three effective tunneling barriers
are pushed in the conduction band and hole
transport takes place throughout the whole structure (c.f. Fig. 2b).
By increasing the voltages on the left and right gate we enter regime (3), which
leads to a significantly decreased current.
In this configuration the Fermi level crosses the two tunneling barriers induced by the two outer constrictions (c.f. Fig. 2c).
Thus the device is in a single quantum dot regime. Finally, if we increase the voltage on the central gate as well, the current is almost completely suppressed [region (1)]. The Fermi level crosses in this configuration all three effective energy gaps and the device is in the DQD regime (c.f. Fig. 1d). The LQD and RQD are consequently defined by the three energy gaps as shown in Figure 1d.
The two slopes separating the different regimes are determined by the
relative lever arms of the lateral gates (CG, LG* and RG) on the three tunneling barriers.
While the two outer tunneling
barriers can be controlled well by the left, right and the central gate ($\alpha=\Delta V_{lg^*,rg}/\Delta V_{cg} \approx$~0.55),
the dominating central barrier can only
be weakly tuned by $V_{lg^*}$ and $V_{rg}$ ($\beta=\Delta V_{cg}/\Delta V_{lg^*,rg} \approx$~0.13), which is
in good agreement with
the geometry of the device.

By fixing the central gate voltage we can record charge stability diagrams in different regimes by varying
the left and right gate voltages independently.
In Figure 2d we show a finite bias ($V_b$ = 10~mV, T = 1.3~K)
measurement taken in regime (1) (see black bar in Fig.~2a).
A honeycomb pattern characteristic for the charge stability diagram of a DQD
can be observed~\cite{wie03}. Electrical transport through the DQD is only
possible in the case where the energy levels in both dots are aligned within the source-drain
bias window. In close analogy to single-layer graphene, the
overall transmission through the DQD is modulated due to local resonances in
the narrow constrictions. This can be seen best in Figure 2e, where
we show a charge stability diagram recorded in regime (3) (c.f. Fig. 2c)
but with lifted left barrier ($V_{lg^*}=0$ and $V_{lg}<0$). Thus the
Fermi level only probes localized states in the right tunneling barrier.
In this measurement we observe only one dominating slope ($\Delta V_{lg}/\Delta V_{rg} \approx$~0.2,
dashed line), which moreover corresponds to the slope of the transmission modulation in
Figure~2d~\cite{com01a}.
In regions of high transmission (e.g. lower left corner in Fig.~2d)
the connecting lines between the triple points become visible. Along these
lines, only one of the dot levels is within the bias window, leading
to current by cotunneling processes.
The relative lever arms between the left and right gates acting on the two dots (LQD and RQD) are
determined from the slopes of these cotunneling lines (see dashed lines in Fig. 2d)
delimiting the hexagons, $\alpha^{R}_{rg,lg} = 0.85$ and $\alpha^{L}_{lg,rg} = 0.12$.
Please note that in this system the RQD can almost equally be tuned with the RG as with
the LG, which again is consistent with the device geometry.

A high-resolution close-up of Figure 2d, but with reversed bias, $V_b$~= -10~mV is shown in Figure 3a.
The bias-dependent extensions of the triangular-shaped regions allows the
determination of the conversion factors between gate voltages and energies (see illustration in Fig. 3b)~\cite{wie03}.
The lever arm between the left (right) gate and the left (right) dot is $\alpha^{L}_{lg} = V_b/\delta V_{lg} = 0.056$ ($\alpha^{R}_{rg} = 0.019$)~\cite{mol09, wie03}.
This allows to extract the single-dot addition energies
$E^L_a=\alpha^{L}_{lg} \cdot \Delta V_{lg}$ = 23~meV and $E^R_a=\alpha^{R}_{rg} \cdot \Delta V_{rg}$ = 13~meV,
which reflect an asymmetry in the QD island sizes.
 The two
quantum dots are considered to be mainly formed by the width-modulated bilayer graphene structure and the
local disorder potential.
In particular, the disorder potential may have significant impact influencing the extend and location of the individual tunneling barriers~\cite{sta09}.
In Figure 3c we show a similar measurement from a different cool down. The BG voltage has been fixed in the center of the transport gap. We observe slightly modified dot sizes with addition energies
of $E^L_a$= 18~meV and $E^R_a$= 21~meV.
Interestingly, the extracted addition energies are
in reasonable agreement with values from single-layer graphene quantum dots with a similar size~\cite{gue10}.
In the following we focus on this more symmetric regime, which
moreover has been investigated in a dilution refrigerator with an electron temperature
of around 100~mK. 
The increased energy resolution makes the edges of the triangles much
more defined, such that the mutual capacitive coupling energy ($E^m$) between the
two QDs can be extracted quantitatively. The coupling energy between both dots can be determined
from the splitting of the triple points, $E^m=\alpha^{L}_{lg} \cdot \Delta V^m_{lg}=\alpha^{R}_{rg} \cdot \Delta V^m_{rg}$=2.4~meV. (see Fig. 3b and arrows in Fig. 3c).

\begin{figure}[tb]\centering
\includegraphics[draft=false,keepaspectratio=true,clip,%
                   width=1.0\linewidth]%
                   {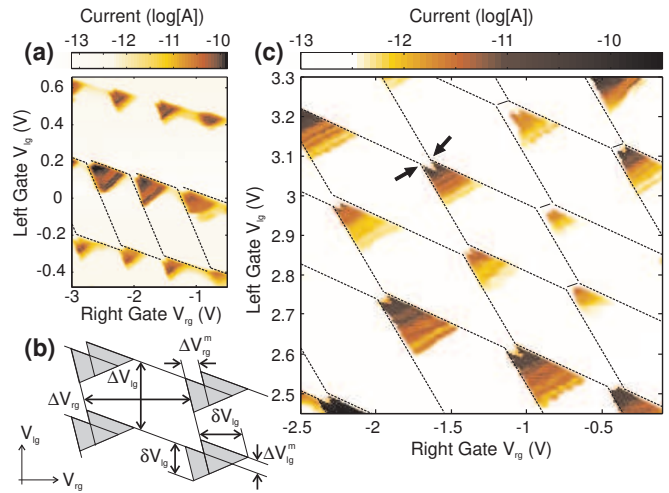}
\caption[FIG3]{
(color online) (a) High-resolution close-up of Figure 2d but with reversed bias ($V_b$=-10~mV; T=1.3~K) highlighting the
double dot characteristic honeycomb pattern.
(b) Schematic charge stability diagram denoting all quantities necessary to deduce the gate lever arms, addition energies and the mutual capacitive coupling energy (for more information see text).
(c) Similar measurement as in panel (a) but at a different cool down and an electron temperature of around 100~mK. The well defined triple points allow to extract the inter-dot coupling energy (see arrows) and additional excited states can be observed.
}
\label{fig3}
\end{figure}

\begin{figure*}[tb]\centering
\includegraphics[draft=false,keepaspectratio=true,clip,%
                   width=0.95\linewidth]%
                   {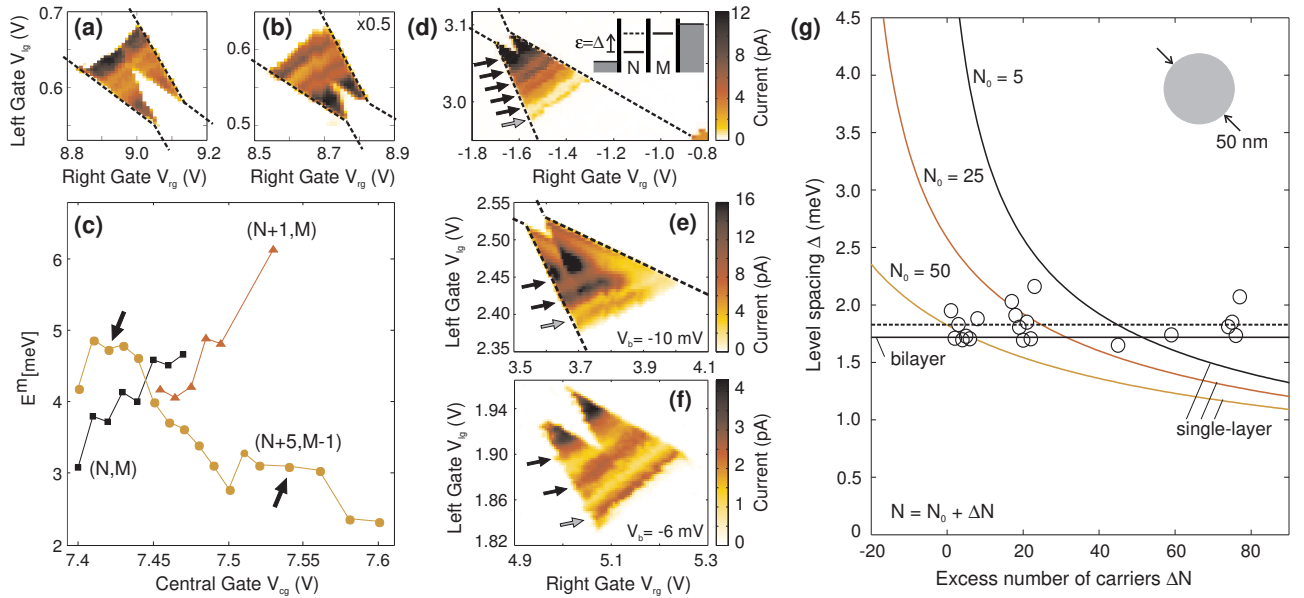}
\caption[FIG4]{
(color online) (a,b) Very same triple points in a weaker ($V_{cg} = 7.54~$V, $V_b = 6~$mV) and stronger ($V_{cg} = 7.42~$V) inter-dot coupling regime respectively.
(c) Mutual capacitive coupling energy $E^m$ depending on the central gate voltage $V_{cg}$, recorded for three individual triple points. N, M denote the number of electrons occupying the dots [see also inset in panel (d)].
(d) High-resolution close-up of Fig. 3c. The triple points show a dense set of electronic excited states. Inset: Schematic energy diagram showing the ground states of the double system and the first excited state of the left dot.
(e,f) Different triple points featuring similar excited state spectra ($V_b = 10~$mV and $6~$mV respectively).
(g) Single-level spacing as a function of the relative change of the carrier occupation number on the left dot. The circles show the experimental data and the dashed horizontal line marks the experimental mean value ($\Delta$=1.82~meV). The three curves represent the model for a single-layer QD for different absolute occupation numbers ($N= N_0 + \Delta N$) while the solid horizontal line marks constant level spacing for a bilayer dot. The QD diameter is assumed to be 50~nm.
}
\label{fig4}
\end{figure*}

By changing the central gate voltage we are able to change the
coupling energy between the left and right QD. Figures 4a and 4b show close-ups of the very same triple point ($V_b$~=~6~mV) for two different inter-dot
coupling energies $E^m$= 4.8~meV ($V_{cg}$=7.42 V, Fig. 4a) and $E^m$= 3.0~meV ($V_{cg}$=7.54 V, Fig. 4b), where
the different triple point separations can be seen.
Figure 4c presents a more detailed analysis of how the coupling energy changes with small variation in $V_{cg}$
for three individual nearby triple points (the examples shown in Figures 4a and 4b are highlighted by arrows).
While we observe  an increase of the coupling
energy for the neighboring triple points [see (N,M) and (N+1,M), where N and M are the number of carriers on the left and right dot, respectively] for increasing $V_{cg}$ we observe a decrease of the coupling
energy by more than a factor of 2 for the triple point (N+5,M-1).
This is in agreement with earlier studies on graphene QDs~\cite{mol09a,liu10} showing a strongly non-monotonic dependence of the conductance on gate voltages due to the
sharp resonances in the constrictions. Interestingly, the extracted coupling energies are roughly by a factor of 2 larger, as compared
to earlier single-layer DQD studies~\cite{mol09a,liu10}, which might be related to the considerably smaller size of the investigated device.

The increased energy resolution also reveals additional
fine structure inside the individual triple points.
This, for
example, can be seen in Figure 3c. The data exhibit distinct lines of
increased conductance parallel to the base line of the triple points.
Figure 4d shows a close-up of Figure 3c highlighting a set of electronic
excited state resonances parallel to the triple points base line (see arrows in
Fig. 4d). Along such a parallel line the inter-dot detuning energy, $\varepsilon$, is
constant and the current is increased due to transport through the excited state (see dashed
lines in the inset in Fig. 4d). The electronic nature of these excited states can be shown by the magnetic
field dependence as discussed below. The detuning energies of these 5 visible excited state resonances are
found to be $\varepsilon$ = 1.7, 3.3, 4.9, 6.5, 8.1~meV, respectively (see black arrows in Fig. 4d). Figures 4e and 4f present high-resolution scans of different triple points recorded
at different bias voltages. The triple points in Figure 4e show excited states parallel to the base line as well as a resonance parallel to its right edge.
The first two excited states are again found at detuning energies of $\varepsilon$ = 1.7 and 3.8~meV (see black arrows).
The conduction suppression near the right edge can be attributed to variations of the coupling between energy levels of the left dot and the source lead~\cite{mol09}.
In Figure 4f we show a triple point pair recorded at $V_b$ = -6~mV with very similar excited state energies ($\varepsilon$ = 1.8 and 3.5~meV; see black arrows).

In total we analyzed more than 50 excited states in a wide energy range. The electronic excited state energy spacing has been found
to be constant over the entire range and a value of $\Delta = 1.75\pm0.27~$meV has
been extracted.
 Interestingly this
 value is in good agreement with the single-particle
confinement energy in disk-like bilayer graphene QDs, which can be estimated by using the density of
states for bilayer graphene $D(E) = \gamma_1 / \pi (\hbar v_F)^2$, where $v_F$ is the
Fermi velocity and $\gamma_1$ = 0.39~meV is the inter-layer hopping energy. Consequently the single-particle level
spacing is given by
\begin{equation}
\Delta = \frac{4 \hbar^2 v_F^2}{\gamma_1} \frac{1}{d^2}, \nonumber
%\notag
\end{equation}
where $d$ is the diameter of the bilayer QD.
Assuming a QD diameter of $d$ = 50~nm, which is
in reasonable agreement with the lithographically defined
QD we obtain a constant (carrier number independent) level
spacing of $\Delta$ = 1.71~meV, which is in good agreement with
the experimental data.

The charge carrier number independent level spacing is in contrast
to single-layer graphene QDs where  
 the
single-particle level spacing, $\Delta(N) = \hbar v_F / d \sqrt{N}$, depends on the number of carriers,
$N$~\cite{sch08}.
In Figure 4g we show the single-level spacing for a 50~nm diameter single-layer
and bilayer graphene QD as function of the excess number of
carriers on the left dot, $\Delta N=N-N_0$.
The solid horizontal line marks the constant level spacing for the bilayer QD, whereas
the three $1/\sqrt{N}$-dependent curves belong to a single-layer QD with different
absolute numbers of charge carriers on the dot ($N_0$ = 5, 25 and 50).
The experimental data, which are depicted by circles in Fig.~4g
have been extracted from charge stability diagrams as shown e.g. in Fig.~3c.
For larger excess carrier numbers ($\Delta N > 40$) we also made use
of relative gate lever arms to estimate $\Delta N$. Since we cannot fully exclude charge
rearragements, these data have to be taken with care.
Please note that the total carrier numbers ($N$, $M$) on the two quantum dots are unknown. In Fig.~4g we plot
only excited states energies extracted from measurements at $V_b<0$ and we
focus on changes of the number of carriers on the left QD only ($\Delta N$, see also inset in Fig.~4d).
For this limited set of excited states we find a constant value of $\Delta=1.82 \pm 0.14$~meV, which is depicted by the dashed horizontal line in Fig.~4g.
Please note that a constant level spacing might also result from an effective constant density of states induced by disorder. However, following Ref.~\cite{gue09}, we estimate the effective dot disorder potential to be on the order of the addition energy of 5-10 charge carriers. Since we are probing a significantly larger energy range we can rule out disorder to be cause of the observed constant level spacing, and attribute it to true band structure properties. Consequently, the model suitable for describing single-layer QDs [$\Delta (N) \propto 1/\sqrt{N}$] does not fit the experimental data for any reasonable $N_0$ whereas the model for a bilayer QD [$\Delta (N) = const.$] is in  good agreement. This leads to the conclusion that the investigated QDs are indeed extending over
both graphene sheets forming the bilayer system.

\begin{figure}[tb]\centering
\includegraphics[draft=false,keepaspectratio=true,clip,%
                   width=0.9\linewidth]%
                   {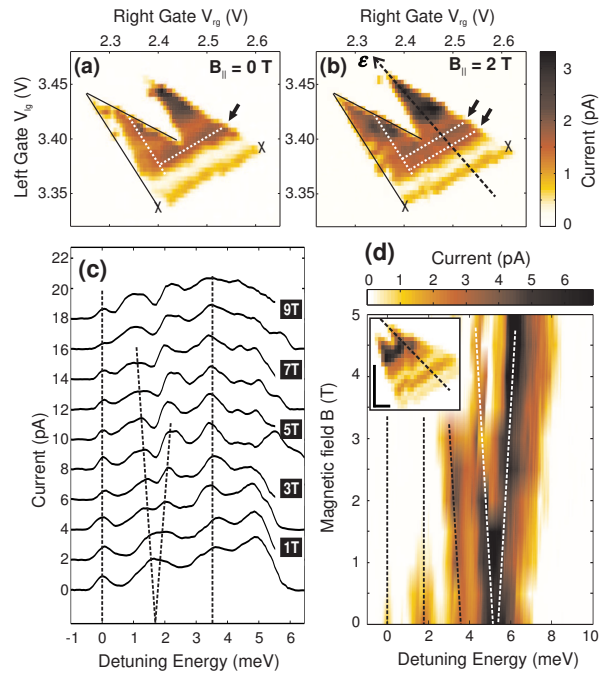}
\caption[FIG5]{
(color online) (a,b) The very same individual triple points recorded at $B_\| = 0~$T and $B_\| = 2~$T respectively ($V_b = -6~$mV).
(c) Current as a function of the detuning energy [measured along the dashed line indicated in panel (b)] for different $B_\|$-fields up to 9~T. Traces are offset by 2~pA for clarity.
(d) Current as a function of the detuning energy and the $B_\|$-field recorded at a different triple point. The corresponding triple point is shown in the inset,
where the scale bars are 50~mV in gate voltages.
}
\label{fig5}
\end{figure}

Figure 5 shows a study of the evolution of excited states as function of the magnetic field oriented parallel to the bilayer graphene plane proving the electronic nature of these states.
In Figures 5a and 5b we show a pair of triple points ($V_b$~=~-6~mV) for two different values of magnetic field.
Excited states parallel to the base line as well as one parallel to the left edge of the triangle are visible (dotted lines). For zero magnetic field again a level spacing of 1.8 meV (see e.g. arrow in Fig. 5a) can be observed. Figure 5b shows the very same triple points at a parallel magnetic field ($B_{\|}$) of 2~T. The position of the triangles in the gate voltage plane almost does not change at all (see crosses and solid lines in Fig. 5a,b). This holds even for $B_{\|}$-fields up to 9~T (as shown in Fig.~5c) and shows that the effects on the orbital parts of the wave functions are unaffected.
The parallel magnetic field has two effects on the triple points, i.e. triangles. First, the conductance at large detuning values slightly increases with increasing
$B_{\|}$-field. This is particularly true for the left triangle, where even parts of the triangle (left edge) are strongly suppressed at low
$B_{\|}$-fields. Since the associated slope differs from the actual triple point edge we attribute this effects to a modulated transmission from the
drain to the right dot ($V_b<0$).
Second, the 1$^{st}$ excited state parallel to the base line splits into two separated peaks (see dotted lines in Fig. 5b). This peak splitting is linear in magnetic field up to roughly 5~T as shown in Figure~5c, where we show current line cuts along the detuning energy axis (see dashed line in Fig.~5b). The different traces are recorded for different parallel magnetic fields (see labels at the right side) and they are offset by 2~pA in current for clarity. We observe that the position of the base line ($\varepsilon$=0) and the 2$^{nd}$ excited state ($\varepsilon$ = 3.6 meV) are constant up to 9~T as function of the $B_{\|}$-field (see vertical dashed lines in Fig. 5c). In contrast, the 1$^{st}$ excited state splits into two peaks, where the characteristic linear peak splitting is 0.2~meV/T (see diverging dashed lines). A similar peak splitting can also be observed on different triple points.
In Figure 5d we show for example
measurements taken on a triple point, where
the 3$^{rd}$ excited state ($\varepsilon$ = 5.3 meV) is splitting.
Here the current at $V_b$~=~-8~mV is plotted as a function of detuning energy and parallel magnetic field. The inset shows the corresponding triple points as well as the detuning axis (dashed line).
While the position of the baseline and the 1$^{st}$ excited state are constant in $B_{\|}$-field, the 2$^{nd}$ excited state shows a down-shift in energy and the 3$^{rd}$ excited state clearly splits linearly in two peaks (see dashed lines).  The base line as well as the 1$^{st}$ and 2$^{nd}$ excited state
completely vanish at increased parallel magnetic fields. The observed B-field dependent peak shifts have again a slope of either 0 or $\pm$ 0.1 meV/T.

All observed peak splittings are likely to be originated from Zeeman spin splitting.
This is motivated by the linear $B_{\|}$-field dependence of the peak splittings (up to high fields) and their characteristic
energy scale,
which is in reasonable agreement with $g \mu_B$~=~0.116~meV/T,
assuming a $g$-factor of 2. The $g$-factor 2, moreover, would be in agreement with recent spin-resolved quantum interference measurements in graphene~\cite{lun09} and observations of spin states in a single-layer graphene quantum dot~\cite{gue10}.
However, a detailed analysis of the different splitting and non-splitting  ($\varepsilon$-dependent) inter-dot transitions is difficult since the carrier number, including the total spin and the degree of lifting the valley degeneracy are not known.
Possible transitions leading to peak splittings may be traced back to increased total spin numbers, non-spin conserving inter-dot tunneling or different g-factors in both quantum dots.
Moreover, we can not exclude that the local $g$-factor is increased by an unintentional layer-dependent doping or by the local gate potentials, which both could potentially lead to an out of plane electric field, which is expected to increase the effective $g$-factor in bilayer graphene~\cite{zha10}.

In summary, we have fabricated and characterized a
tunable bilayer graphene double quantum dot based on a
width-modulated graphene nanostructure with lateral
graphene gates.
Its functionality was demonstrated by the observation of a tunable inter-dot coupling energy and the
electronic excited state spectra over a wide energy range.
We have shown that - in contrast to single-layer graphene - the single-particle level spacing is independent of the number of charge carriers on the bilayer quantum dots. By applying a $B_{\|}$-field we observed inter-dot transition energy splittings on the order of Zeeman splittings.
These results
give insights into tunable bilayer graphene double quantum dot
devices and open the way to study individual spin states and spin
coherence times in more detail in future experiments.

{Acknowledgment ---}
We thank A. Steffen, R. Lehmann and U. Wichmann for the help on sample fabrication and electronics.
Discussions with H.~L\"uth, J. G\"uttinger, R.~Leturcq, D.~DiVincenzo, T.~Ihn, S. Dr\"oscher, F.~Libisch, F.~Haupt, H.~Bluhm, G.~Burkard and D.~Schuricht
and support by the JARA Seed Fund and the DFG (SPP-1459 and FOR-912) are agratefully
acknowledged.


\begin{thebibliography}{99}
\bibitem{com01a}
Figures 2d and 2e have the same scaling aspect ratio.
\end{thebibliography}
\end{document}